\documentclass{iopjournal}

\usepackage[super,sort&compress,comma]{natbib}
\usepackage{lmodern}
\usepackage[final]{microtype}  
\usepackage{hyperref}
\usepackage{float} 
\usepackage{ragged2e} 
\usepackage{longtable}
\usepackage{booktabs}
\begin{document}

\articletype{Article type} 


\title{Sensitivity of silicon-to-water dose conversion and Bragg-peak metrics to stopping-power datasets in proton dosimetry}


\author{F. Matias$^{1,*}$\orcid{0000-0002-9339-8286}, J. M. B. Shorto$^1$\orcid{0000-0003-4817-5275}, P. de Vera$^2$\orcid{0000-0002-5645-412X}, R. Garcia-Molina$^2$\orcid{0000-0001-8755-8709}, I. Abril$^3$\orcid{0000-0001-9366-3484}, T. F. Silva$^4$\orcid{0000-0002-7643-2198}, J. Pereira$^{1,5}$\orcid{0000-0002-0755-8774} and H. Yoriyaz$^{1,*}$\orcid{0000-0003-4023-236X}}

\affil{$^1$Instituto de Pesquisas Energéticas e Nucleares, Avenida Professor Lineu Prestes, São Paulo, 05508-000, Brazil}

\affil{$^2$Departamento de Física, Centro de Investigación en Óptica y Nanofísica, Universidad de Murcia, Murcia, 30100, Spain}

\affil{$^3$Departament de Física, Universitat d’Alacant, Alacant, 03080, Spain}

\affil{$^4$Instituto de Física da Universidade de São Paulo, Rua do Matão, travessa R187, São Paulo, 05508-090, Brazil}

\affil{$^5$National Physical Laboratory, Teddington, TW11 0LW, United Kingdom}

\affil{$^*$Author to whom any correspondence should be addressed.}

\email{phdflaviomatias@gmail.com and yoriyaz@usp.br}

\justifying

\begin{abstract}
\noindent
\textit{Objective.}
Accurate proton dosimetry requires internally consistent stopping-power data for both detector-to-water dose conversion and Monte Carlo radiation transport simulations. This study quantifies the sensitivity of water-to-silicon stopping-power ratios, converted dose-to-water distributions, and Bragg-peak metrics to the stopping-power dataset for a $67.5$ MeV pristine proton beam in water. \textit{Approach.}
PHITS simulations were performed using dataset-specific stopping powers from SRIM-2013, PSTAR/NIST, ATIMA, and TDDFT-Penn for proton energies up to $10$ MeV, combined with a common SBETHE-based high-energy extension above $10$ MeV. Normalized percentage depth-ionization data measured with a PTW silicon diode were converted into relative percentage depth-dose distributions in water using depth-dependent water-to-silicon stopping power ratios. Direct dose-to-water simulations were also performed to evaluate Bragg-peak and distal-range metrics. \textit{Main results.}
The water-to-silicon stopping-power ratio exhibited a pronounced depth dependence, with relative entrance-to-distal variations ranging from $17.40$\% to $22.10$\% across the datasets. All converted relative PDD curves reached their maximum at $36.86$ mm, with no dataset-dependent shift. This position agrees with the experimental Bragg-peak depth of $(36.81 \pm 0.15)$ mm, defined as the midpoint between adjacent PDI maxima at $36.76$ and $36.86$ mm. The converted curves were approximately $3$\% below the reference curve throughout the entrance and plateau regions, while dataset-dependent deviations of up to $2.1$\% were observed near the Bragg peak. Direct PHITS calculations predicted Bragg-peak depths ranging from $36.575$ to $36.675$ mm; among the evaluated options, only the TDDFT-Penn result lay within the quoted experimental uncertainty. \textit{Significance.}
For this benchmark, the adopted stopping-power treatment affects the normalization of the converted dose throughout the entrance, plateau, and distal regions more strongly than it affects the Bragg-peak position. Harmonizing stopping-power data between detector-to-water conversion and Monte Carlo transport, or explicitly including dataset dependence in the uncertainty budget, is therefore important for proton-beam calibration, commissioning, quality assurance, and Monte Carlo validation.

\end{abstract}

\section{Introduction}

Precise determination of the absorbed dose in clinical proton beams is essential for commissioning proton therapy systems, routine quality assurance (QA), and verification and benchmarking of Monte Carlo (MC) dose calculations. These activities are supported by widely adopted codes of practice and recent community guidelines for the acquisition of proton beam data and reference dosimetry \cite{Goma:2024,Green:2025,IAEA:2024}. Here, QA workflows refer to standardized measurement procedures used to ensure consistent beam delivery within predefined tolerances. 

Recent work from national metrology institutes, including the UK National Physical Laboratory (NPL), has further consolidated the absorbed-dose-to-water framework for clinical proton beams \cite{Palmans:2022,Lourenco:2022,Lourenco_2023}. In practical applications, depth-dose and related beam-quality measurements are frequently performed with solid-state silicon-based detectors (e.g., diodes and pixel detectors) as part of beam characterization and QA workflows. The measured signal is commonly converted to water-equivalent quantities using appropriate stopping-power ratios (SPRs) \cite{Hamad:2025}. Although such conversions can yield accurate results for pristine Bragg peaks when the local proton energy spectrum remains comparatively narrow, their sensitivity to spectral variations increases with depth as the energy distribution broadens in modulated beams~\cite{BIANCHI:2023,BORTOT:2024,Laitano:2000,IAEA:2024}. In this context, SPR-related uncertainties remain a major contributor to range uncertainty, motivating continued efforts to improve SPR prediction and control systematic effects \cite{Huijskens:2025}. In parallel, MC transport codes may rely on different stopping-power compilations and physics implementations, and experimental validations have shown that these choices can lead to measurable differences in SPR-related quantities \cite{Liu:2022}.

In this work, we analyze a reference 67.5 MeV pristine proton beam in water using a published benchmark geometry~\cite{Faddegon:2015}. The choice of this specific pristine beam was deliberate: as a well-established benchmark in the literature, it enables a controlled assessment of stopping-power sensitivity while avoiding the additional spectral complexity inherent in modulated spread-out Bragg peak (SOBP) fields. We address two related questions: (i) how the choice of stopping-power dataset affects the conversion of silicon depth-ionization data into relative dose-to-water distributions via water-to-silicon stopping-power ratios, and (ii) how it affects the Bragg-peak depth and range extracted from depth-dose curves determined directly in water. To this end, we consider four stopping power datasets for water and silicon: PSTAR/NIST \cite{PSTAR:2005}, SRIM-2013 \cite{ZIEGLER:2010}, ATIMA stopping powers as used internally by PHITS \cite{PHITS:2024,Matsuya:2024,PHITS:2025,ATIMA,Weick:2018}, and TDDFT-Penn stopping powers provided as user tables \cite{MATIAS:2024,MATIAS-JCP:2024,MATIAS-PRL:2025}. All datasets are implemented in otherwise identical PHITS simulations to isolate the stopping-power sensitivity.


\section{Methods}

\subsection{Monte Carlo modeling and stopping-power ratio formalism for dose conversion}

\begin{figure}[H]
\centering
\includegraphics[width=0.58\textwidth]{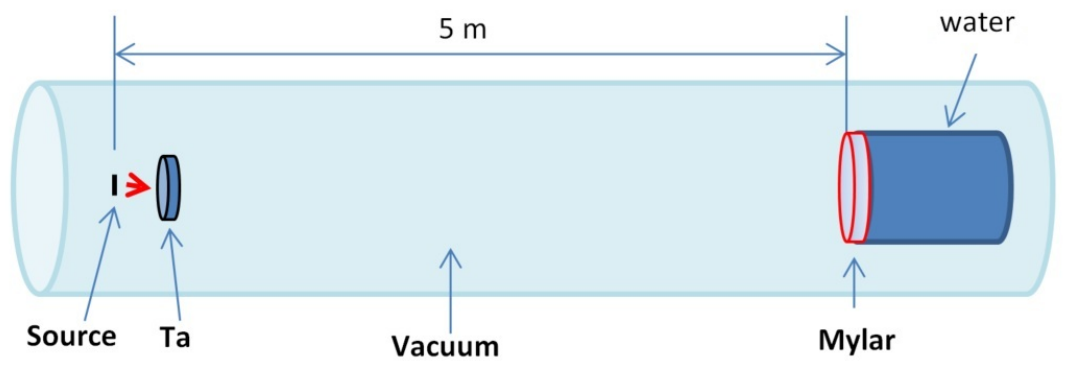}
\caption{Schematic geometry of the experimental setup for determining proton fluence spectra and depth-dose benchmarking.}
\label{fig1}
\end{figure}

The transport of protons in water was modeled using the PHITS MC (v3.35) code \cite{PHITS:2024,Matsuya:2024,PHITS:2025} to determine the proton fluence spectra as a function of energy in a cylindrical water phantom. The setup is described by Faddegon \emph{et al.}~\cite{Faddegon:2015}: a proton beam of energy $67.5$ MeV with an energy spread of $0.4$ MeV and an oval field size of $5\times 10$ mm$^2$ traverses a thin tantalum (Ta) scattering foil, propagates through a vacuum drift of approximately $5$ m, and subsequently passes through a Mylar entrance window before entering the water phantom (see Fig.~\ref{fig1}). Many other components, such as scattering foil plates, collimators, Kapton, and plugs, have been included in the model but omitted from the figure. Thin detection regions were placed at the detector depth positions used in the experiment. Unless otherwise stated, the maximum statistical uncertainty in the MC simulations was kept below $1$\%.

PHITS~\cite{PHITS:2024,Matsuya:2024,PHITS:2025} employs the ATIMA~\cite{ATIMA,Weick:2018} algorithm as its default method for stopping-power calculations and also supports user-defined stopping-power tables. We used this capability to run otherwise identical simulations with alternative stopping-power datasets for water and silicon (PSTAR/NIST~\cite{PSTAR:2005}, SRIM-2013~\cite{ZIEGLER:2010}, and TDDFT-Penn~\cite{MATIAS:2024,MATIAS-JCP:2024,MATIAS-PRL:2025} tables). Depth-dose profiles were obtained with PHITS by modeling a water phantom divided into 1000 axial layers, each 0.05 mm thick, and computing the absorbed dose in every layer to generate the depth-dose distribution. 
The position of the dose maximum (Bragg peak) and the distal $80$\% proton range, \(R_{80}\), were subsequently determined. Here, \(R_{80}\) is defined as the depth, measured beyond the Bragg peak, at which the absorbed dose falls to $80$\% of its maximum value.

Within the Bragg-Gray cavity formalism, the absorbed dose in water at depth $z$ can be determined by scaling the dose deposited in the detector's silicon sensitive volume, in the case of a diode, by the corresponding water-to-silicon SPR.~\cite{Faddegon:2015,IAEA:2024} Consequently, converting the raw detector response, or the dose measured in silicon, into an absorbed dose distribution in water requires applying an appropriate water-to-silicon SPR.

For a proton fluence spectrum at depth $z$, the dose-to-water conversion can be written in its fluence-weighted form as~\cite{IAEA:2024,Palmans:2022,Lourenco:2022,Lourenco_2023}

\begin{equation}
D_{w}(z)
=
D_{\mathrm{Si}}(z)
\frac{
\displaystyle \int_{0}^{E_{\max}}
\Phi_{w}(z,E)\,S_{w}(E)\,dE
}{
\displaystyle \int_{0}^{E_{\max}}
\Phi_{\mathrm{Si}}(z,E)\,S_{\mathrm{Si}}(E)\,dE
},
\label{eq:dose_fluence_weighted}
\end{equation}
where $D_{w}(z)$ is the absorbed dose in water at depth $z$, $D_{\mathrm{Si}}(z)$ is the absorbed dose in silicon at the same depth, $\Phi_{w}(z,E)$ and $\Phi_{\mathrm{Si}}(z,E)$ are the energy-dependent proton fluence spectra in water and silicon, respectively, and $S_{w}(E)$ and $S_{\mathrm{Si}}(E)$ are the corresponding mass stopping powers, i.e., stopping powers per unit density, for water and silicon. The integration is performed over the full energy interval of the proton fluence spectrum up to $E_{\max}$. The energy-dependent proton fluence was calculated using the PHITS T-Track tally, a track-length estimator that scores the length of each particle track segment within the scoring volume.

Eq.~(\ref{eq:dose_fluence_weighted}) can be rewritten in terms of the fluence-weighted mass stopping power as
\begin{equation}
D_{w}(z)
=
D_{\mathrm{Si}}(z)
\frac{
\left(\overline{S}/\rho\right)_{w}(z)
}{
\left(\overline{S}/\rho\right)_{\mathrm{Si}}(z)
}
=
D_{\mathrm{Si}}(z)\,\mathrm{SPR}_{w,\mathrm{Si}}(z),
\label{eq:dose_spr}
\end{equation}
where $\left(\overline{S}/\rho\right)_{w}(z)$ and $\left(\overline{S}/\rho\right)_{\mathrm{Si}}(z)$ are the fluence-weighted mass stopping powers of water and silicon, respectively, and
\begin{equation}
\mathrm{SPR}_{w,\mathrm{Si}}(z)
=
\frac{
\left(\overline{S}/\rho\right)_{w}(z)
}{
\left(\overline{S}/\rho\right)_{\mathrm{Si}}(z)
}
\label{eq:spr}
\end{equation}
is the water-to-silicon stopping power ratio at depth $z$. \cite{Faddegon:2015,IAEA:2024} In this work, the stopping powers used in the SPR were taken from the selected dataset under evaluation, thereby allowing the sensitivity of the converted dose to different stopping power compilations to be quantified.

In practice, the reference data provided by Faddegon \emph{et al}.~\cite{Faddegon:2015} correspond to a normalized percentage depth-ionization (PDI) curve rather than an absolute dose-to-silicon distribution. The present conversion analysis was intentionally restricted to the PDI dataset measured with the PTW silicon diode, for which the experimental distal 80\% range is $R_{80}=(37.35\pm0.15)$ mm. The PTW and EFD diode results were not averaged because the objective of this work is to isolate the influence of the stopping-power ratio on the conversion of a fixed detector response. Using a single detector dataset avoids conflating SPR-dependent effects with differences associated with detector type or response. Therefore, after computing the depth-dependent water-to-silicon stopping-power ratio for each stopping-power dataset, the corresponding relative percentage depth-dose (PDD) curve was obtained as
\begin{equation}
\mathrm{PDD}^{(k)}(z)
=
\mathrm{PDI}(z)\,
\mathrm{SPR}^{(k)}_{w,\mathrm{Si}}(z),
\label{eq:pdi_to_pdd}
\end{equation}
where $k$ denotes the stopping-power dataset. The original PTW PDI curve provided by Faddegon \emph{et al}.~\cite{Faddegon:2015} was normalized to 100\% at the Bragg peak. Owing to the discrete experimental depth sampling, two adjacent points have the same maximum value: 100\% at 36.76 mm and 100\% at 36.86 mm. The representative experimental Bragg-peak depth was therefore defined as the midpoint between these two positions, the adjacent PDI maxima at 36.76 and 36.86 mm, with the experimental depth uncertainty retained as $\pm0.15$ mm. This initial normalization was preserved after applying the depth-dependent stopping-power ratio, and no subsequent renormalization was performed. Thus, the converted curves should be interpreted as relative dose-to-water distributions, suitable for assessing dataset-dependent changes in the shape and magnitude of the converted dose, rather than as absolute absorbed-dose calibrations.

\begin{figure}[H]
\centering
\includegraphics[width=0.85\textwidth]{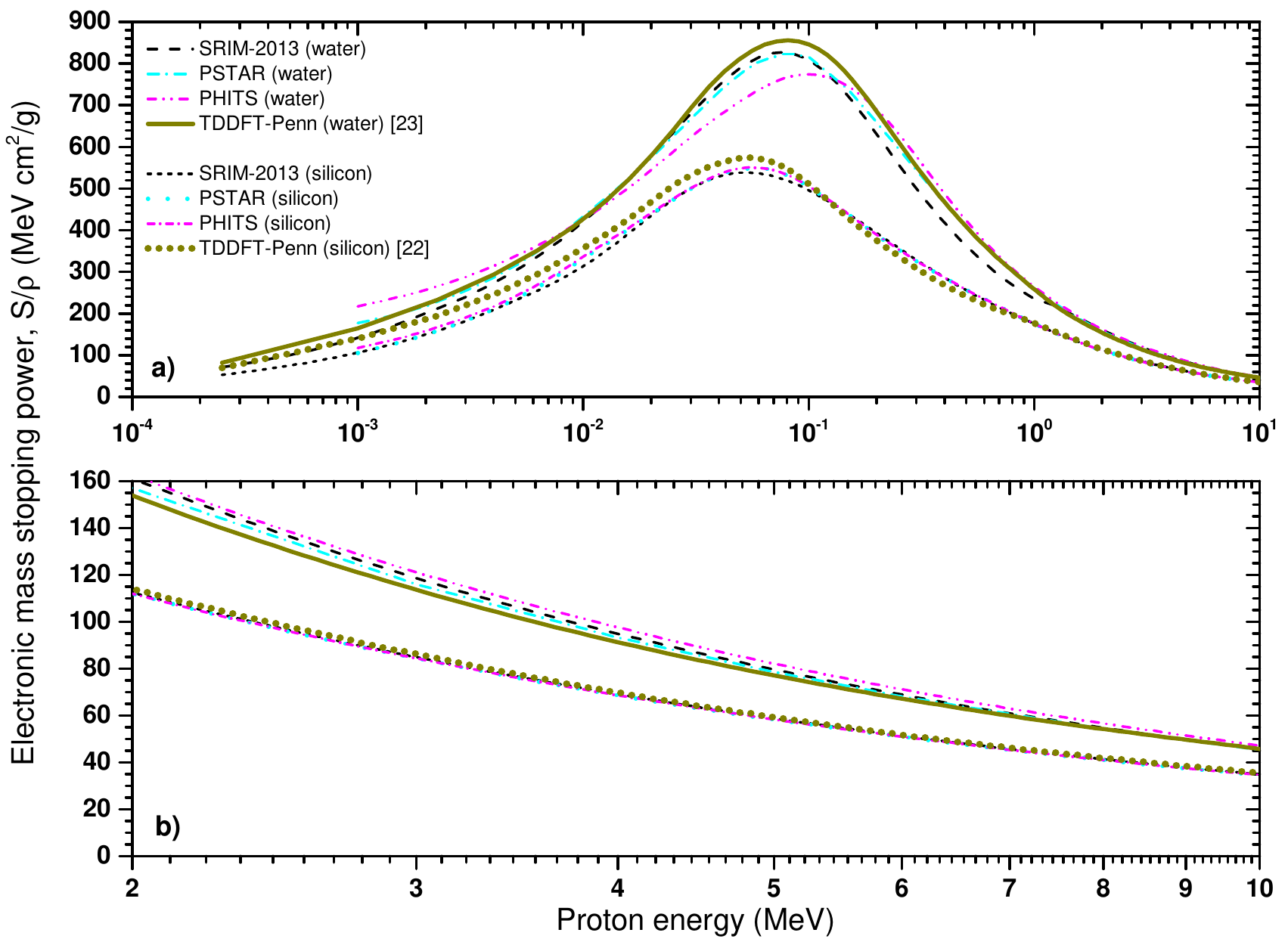}
\caption{
Proton electronic mass stopping powers, $S/\rho$, for water and silicon from SRIM-2013, PSTAR/NIST, PHITS internal ATIMA, and TDDFT-Penn~\cite{MATIAS-PRL:2025,MATIAS-JCP:2024} datasets. Panel (a) presents the broad low- to intermediate-energy range extending up to $10$ MeV, whereas panel (b) provides an enlarged view of the $2-10$ MeV sub-interval, corresponding to the region probed in proximity to the distal side of the Bragg peak. Dataset-dependent differences propagate into $\mathrm{SPR}_{w,\mathrm{Si}}(z)$ and therefore into the silicon-to-water dose conversion.
}
\label{fig2}
\end{figure}

Figure~\ref{fig2} compares the mass stopping powers of protons in water and silicon, $S_{w}(E)$ and $S_{\mathrm{Si}}(E)$, for the four stopping power options considered in this study. These datasets constitute the primary physics input to the stopping-power ratio, Eq.~(\ref{eq:spr}), which, in the present case, drives the PDI-to-PDD conversion in Eq.~(\ref{eq:pdi_to_pdd}). Consequently, any systematic difference between $S_{w}(E)$ and $S_{\mathrm{Si}}(E)$ directly translates into a depth-dependent spread of $\mathrm{SPR}_{w,\mathrm{Si}}(z)$ and, therefore, into changes in the converted relative dose magnitude. In direct transport calculations, the same stopping-power differences may also affect Bragg-peak and range metrics. The highest sensitivity is expected in the intermediate-energy interval sampled near the distal side of the Bragg peak, where small relative deviations in stopping power can accumulate into measurable range shifts.


For the stopping-power tables employed in the present simulations, the low- and intermediate-energy domains were treated in a dataset-specific manner. For proton kinetic energies $E \leq 10$ MeV, stopping powers were taken directly from the corresponding models or compilations shown in Fig.~\ref{fig2}. For $E>10$ MeV, the tables were extended using the relativistic SBETHE code, which computes electronic stopping powers from the corrected Bethe formula incorporating shell, density-effect, Barkas, and Lindhard-S{\o}rensen corrections~\cite{Salvat:2023SBETHE,SalvatAndreo:2023SBETHEdata}. This high-energy approach requires the mean excitation energy, $I$, as the principal material-dependent input parameter. 

For liquid water, we adopted $I = 79.4$ eV, consistent with dielectric-response descriptions of liquid water that are constrained by the $f$-sum rule and with recent TDDFT-Penn-based stopping-power calculations~\cite{GarciaMolina:2011PMB,GarciaMolina:2013NIMB,deVera:2019EPJD,MATIAS-PRL:2025}. This choice is further supported by recent studies demonstrating that it minimizes discrepancies in Bragg-peak positioning relative to accurately measured depth-dose distributions~\cite{Kumazaki:2007,Schardt:2007}. The adopted value also lies within the uncertainty interval of the experimental estimate $I=(79.7\pm0.5)$ eV reported by Bichsel and Hiraoka~\cite{BISCHEL1992345}. For silicon, we used $I = 170$ eV, obtained from the $f$-sum-rule-constrained optical response previously employed in our stopping-power calculations~\cite{MATIAS-JCP:2024}.

Consequently, the dataset labels shown in Fig.~\ref{fig2} identify the stopping-power source used up to 10 MeV, where dataset-specific differences are most relevant for the distal Bragg-peak region, whereas a unified SBETHE-based high-energy extension was applied above this energy threshold.

The curves denoted as ``PHITS (water)'' and ``PHITS (silicon)'' in Fig.~\ref{fig2} represent the stopping-power data used internally by PHITS for dose calculations, generated by the ATIMA module. In contrast, the remaining options are implemented using external stopping-power tables to disentangle and quantify the sensitivity of the calculated results to the choice of stopping-power dataset, while keeping all other transport conditions unchanged.

\subsection{Depth positions, mean energies, and residual range}

The energy-dependent proton fluence spectra were calculated at the experimental depth positions used in the benchmark of Faddegon \emph{et al.}~\cite{Faddegon:2015}. At each depth $z$, the mean proton energy $\bar{E}(z)$ was obtained from the corresponding energy-dependent proton fluence spectrum. To characterize the local beam quality along the depth-dose curve, we also report the continuous slowing down approximation (CSDA)-equivalent residual range in water associated with $\bar{E}(z)$. This quantity represents the remaining range of a proton with energy $\bar{E}(z)$ and should not be confused with the distal dose-based range metric $R_{80}$ extracted from the depth-dose distributions.

The resulting depth positions, mean proton energies, and CSDA-equivalent residual ranges are listed in Table~\ref{tab:depth_energy}. These quantities are used to identify the energy interval sampled at each measurement depth and to support the interpretation of the depth-dependent water-to-silicon stopping-power ratio.


\begin{table}[H]
\caption{Depth positions, mean proton energies, and CSDA-equivalent residual ranges in water used in this work. The mean energies were obtained from the calculated proton fluence spectra at each depth, whereas the residual ranges correspond to the remaining CSDA range in water associated with those mean energies.}
\centering
\setlength{\tabcolsep}{7pt}
\begin{tabular}{r r r}
\hline
Depth (mm)~\cite{Faddegon:2015} & Mean energy (MeV) & Residual range (mm)\\
\hline
1.06 & 64.275 & 35.00 \\
6.06 & 63.461 & 34.22 \\
11.06 & 54.579 & 26.05 \\
16.06 & 51.193 & 23.28 \\
21.06 & 41.804 & 16.16 \\
26.06 & 33.982 & 11.12 \\
31.06 & 23.588 &  5.77 \\
36.06 &  8.195 &  0.86 \\
36.26 &  7.638 &  0.76 \\
36.36 &  7.619 &  0.76 \\
36.46 &  7.061 &  0.66 \\
36.56 &  6.766 &  0.62 \\
36.66 &  6.696 &  0.61 \\
36.76 &  6.166 &  0.52 \\
36.86 &  6.087 &  0.51 \\
36.96 &  5.562 &  0.44 \\
37.06 &  5.264 &  0.40 \\
37.16 &  4.973 &  0.36 \\
37.26 &  4.890 &  0.35 \\
37.36 &  4.420 &  0.29 \\
37.56 &  4.083 &  0.26 \\
37.81 &  3.375 &  0.18 \\
38.06 &  2.928 &  0.14 \\
38.31 &  2.671 &  0.12 \\
38.56 &  2.944 &  0.15 \\\hline
\end{tabular}
\label{tab:depth_energy}
\end{table}

\section{Results}

\subsection{Mean energy and residual range versus depth}

\begin{figure}[H]
\centering
\includegraphics[width=0.58\textwidth]{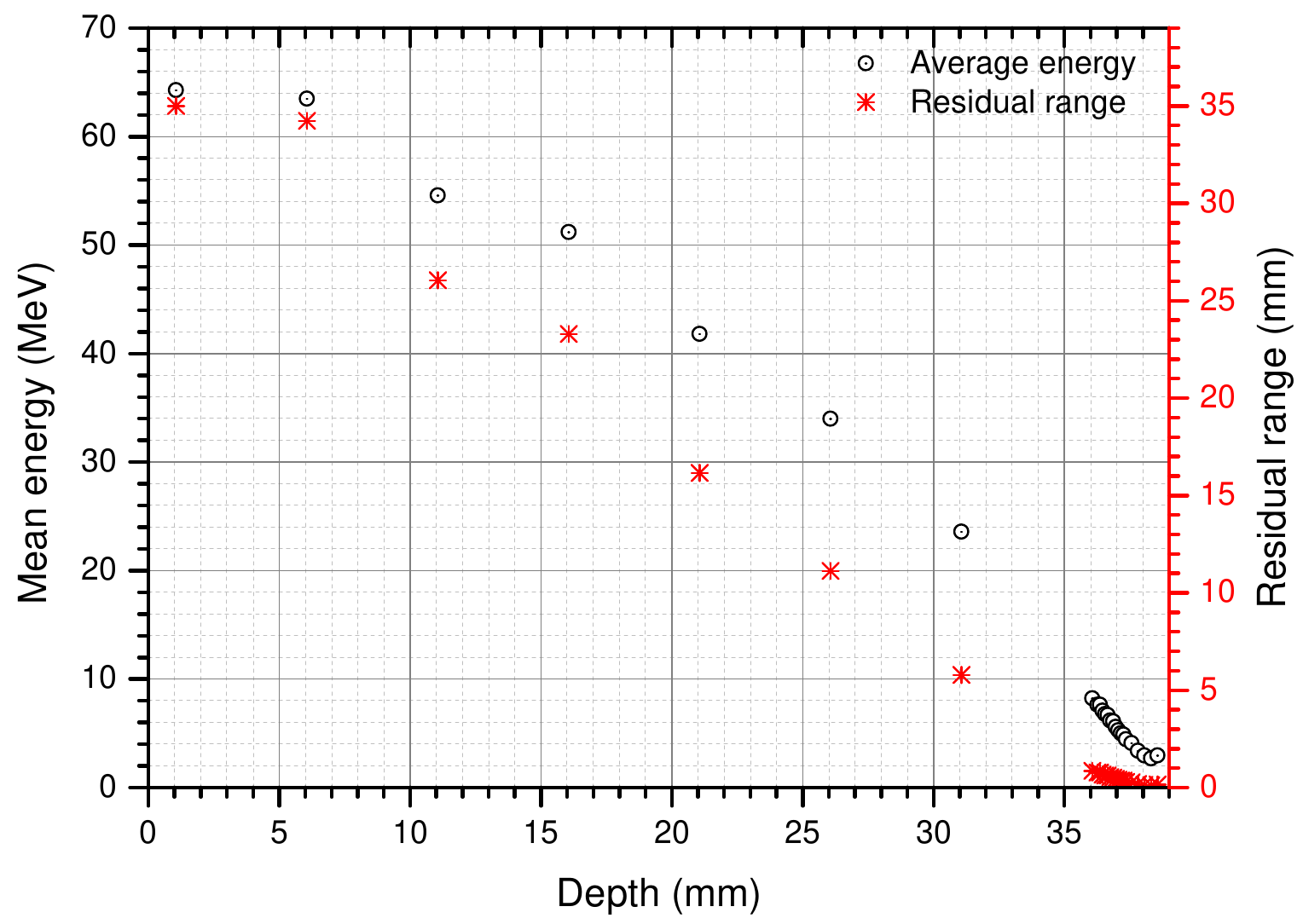}
\caption{
Mean proton energy and corresponding residual range as a function of depth in water for the measurement positions used in this work. The values were derived from the proton fluence spectra calculated at each depth. The residual range decreases toward the distal region of the Bragg peak, where the proton spectrum is sampled in the low- to intermediate-energy interval relevant for the depth-dependent water-to-silicon stopping-power ratio.
}
\label{fig3}
\end{figure}

Figure~\ref{fig3} summarizes the mean proton energies and the corresponding residual ranges derived from the proton fluence spectra at each measurement depth (see Table~\ref{tab:depth_energy}). Both quantities decrease overall as the beam approaches the distal region of the Bragg peak. The benchmark depths include plateau positions from $1.06$ to $31.06$ mm and a dense sampling of the Bragg-peak region from $36.06$ to $38.56$ mm, where the mean proton energy lies in the low- to intermediate-energy interval. This region is particularly relevant for evaluating the depth dependence of the water-to-silicon stopping-power ratio.


\subsection{Depth-dependent SPR variation}

The depth-dependent SPR variation reported in Table~\ref{tab:spr_variation} was quantified as the relative change between the entrance depth, $z=1.06$ mm, and the deepest sampled depth, $z=38.56$ mm:
\begin{equation}
\Delta \mathrm{SPR}^{(k)}
=
100\,
\frac{
\mathrm{SPR}^{(k)}_{\mathrm{w,Si}}(38.56~\mathrm{mm})
-
\mathrm{SPR}^{(k)}_{\mathrm{w,Si}}(1.06~\mathrm{mm})
}{
\mathrm{SPR}^{(k)}_{\mathrm{w,Si}}(1.06~\mathrm{mm})
}.
\label{eq:spr_variation}
\end{equation}
Here, $k$ denotes the stopping-power dataset used to compute the depth-dependent water-to-silicon stopping-power ratio. Table~\ref{tab:spr_variation} summarizes the resulting variations in SPRs across the different stopping-power parameterization options.

\begin{table}[H]
\caption{Depth-dependent variation of the water-to-silicon SPR between $1.06$ mm and $38.56$ mm depth in water for the different stopping-power options.}
\centering
\renewcommand{\arraystretch}{1.15}
\begin{tabular}{l r}
\hline
Dataset option & SPR variation (\%)\\
\hline
PHITS $\leftarrow$ SRIM-2013   & 17.40\\
PHITS $\leftarrow$ PSTAR (ICRU-90)  & 22.10\\
PHITS $\leftarrow$ ATIMA  & 20.70\\
PHITS $\leftarrow$ TDDFT-Penn  & 20.97\\
\hline
\end{tabular}
\label{tab:spr_variation}
\end{table}

The variation in SPR depends strongly on the stopping-power dataset: PHITS $\leftarrow$ SRIM yields the smallest depth-dependent change ($17.40$\%), whereas PHITS $\leftarrow$ PSTAR yields the largest ($22.10$\%). This systematic spread reflects differences in the relative stopping powers of water and silicon across the relevant energy interval, particularly in the $\sim 2$ to $10$ MeV range (see Fig.~\ref{fig2}), where the stopping power curves diverge more clearly.

\subsection{Dose-to-water conversion from silicon depth-ionization data}

The normalized experimental PDI curve measured with the PTW silicon diode was provided by Faddegon \emph{et al}.~\cite{Faddegon:2015}. For each stopping-power dataset considered in this work, this PDI curve was converted into a relative PDD distribution in water by multiplying the PDI value at each depth by the corresponding depth-dependent water-to-silicon stopping-power ratio, $\mathrm{SPR}_{w,\mathrm{Si}}(z)$, according to Eq.~(\ref{eq:pdi_to_pdd}). No additional renormalization was applied after this conversion.

\begin{figure}[H]
\centering
\includegraphics[width=0.95\textwidth]{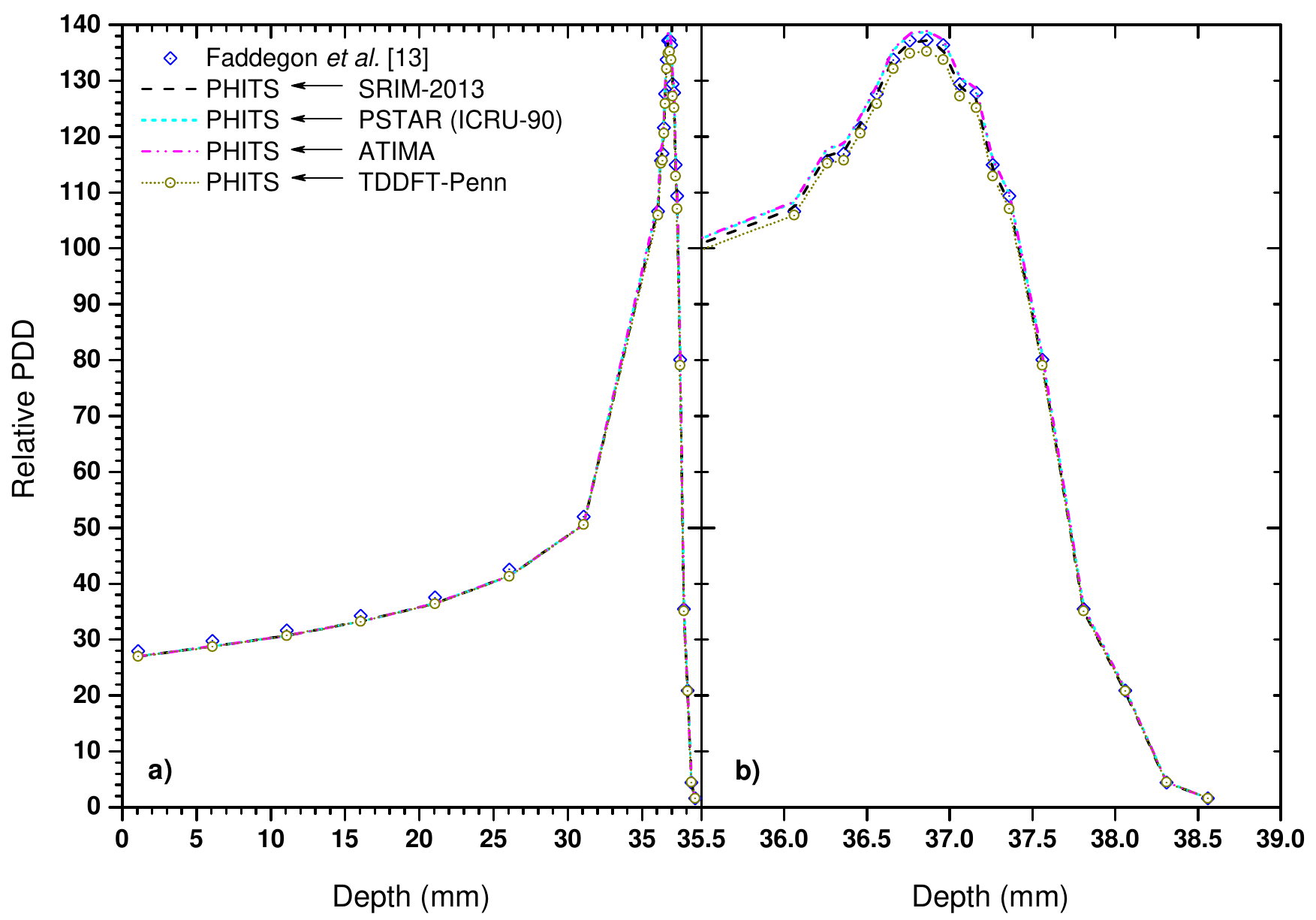}
\caption{Relative PDD distributions in water, derived from the normalized PTW silicon-diode PDI curve provided by Faddegon \emph{et al}.~\cite{Faddegon:2015}, using the depth-dependent water-to-silicon stopping-power ratio, $\mathrm{SPR}_{w,\mathrm{Si}}(z)$. The resulting curves were computed using stopping-power data from SRIM-2013, PSTAR (ICRU-90), the PHITS internal ATIMA implementation, and TDDFT-Penn. Panel (a) presents the complete relative depth-dose distribution, whereas panel (b) shows an enlarged view of the distal fall-off and Bragg-peak regions. All converted curves reach their maximum at $36.86$ mm.}
\label{fig4}
\end{figure}

Figure~\ref{fig4} presents the resulting relative PDD curves. For all stopping-power datasets, the converted PDD reaches its maximum at $z=36.86$ mm. This depth corresponds to one of the two adjacent experimental PDI maxima and is only $0.05$ mm distal to the representative experimental Bragg-peak depth of $(36.81\pm0.15)$ mm, defined as the midpoint of the two maxima. Therefore, the conversion does not introduce a dataset-dependent displacement of the Bragg-peak position within the experimental depth resolution. In contrast, the magnitude of the converted dose near the distal edge depends on the stopping-power dataset, with pointwise discrepancies of up to $2.1$\% around the Bragg peak.



\begin{figure}[H]
\centering
\includegraphics[width=0.85\textwidth]{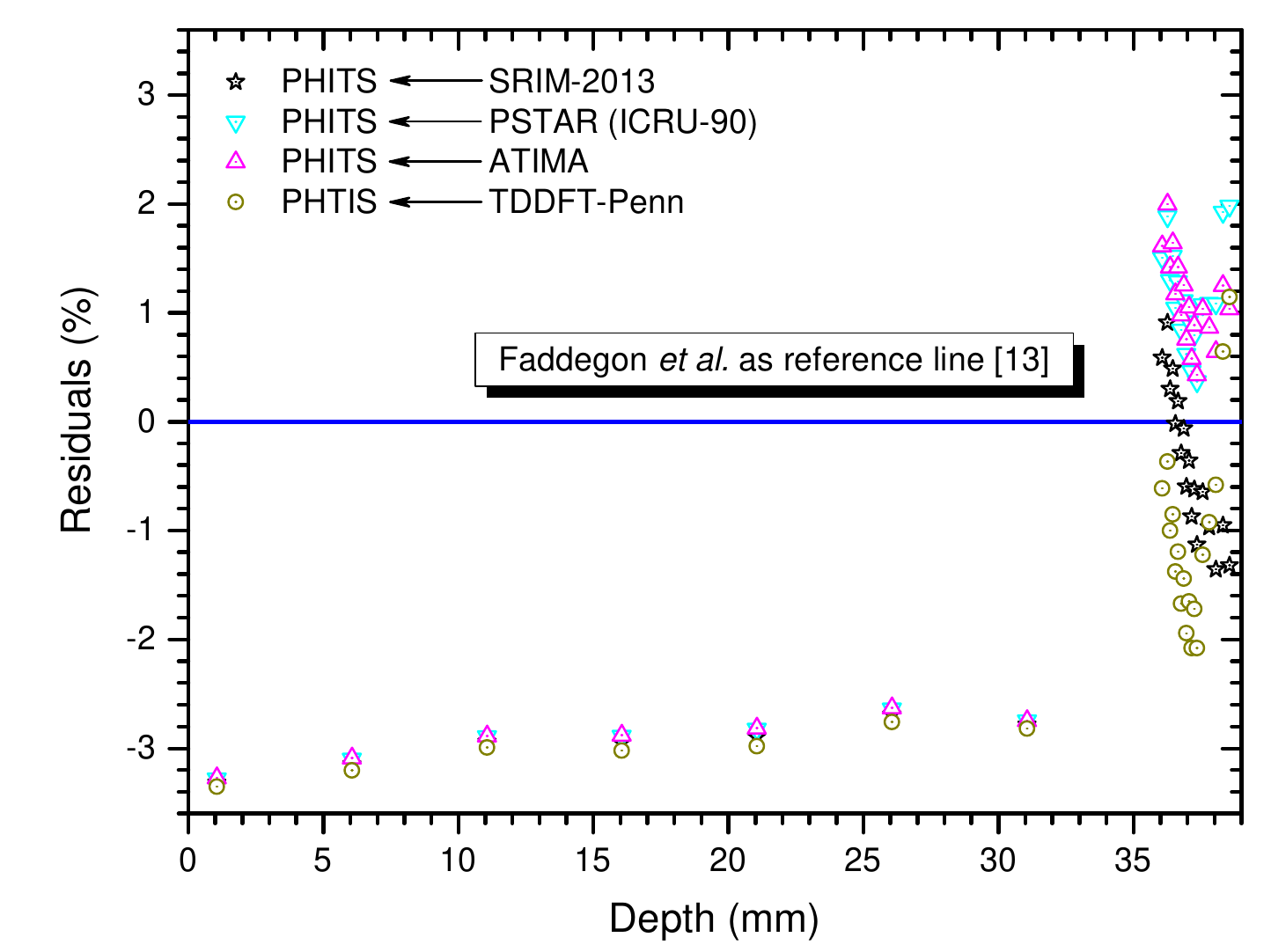}
\caption{Percentage residuals between the relative PDD curves derived from the normalized PTW silicon-diode PDI data provided by Faddegon \emph{et al}.~\cite{Faddegon:2015} and the corresponding reference curve. For each stopping-power dataset, the residuals were computed as the relative deviation from the reference PDD. A systematic offset of approximately $-3\%$ is observed across the entrance and plateau regions, whereas a dataset-dependent dispersion of up to approximately $2.1$\% is evident near the Bragg peak. The horizontal line at zero represents perfect agreement with the reference.}
\label{fig5}
\end{figure}

The close-up view in Fig.~\ref{fig5} shows these deviations in the distal region, where the proton spectrum is dominated by average energies between $3$ and $8$ MeV (see Fig.~\ref{fig3}). In this regime, small differences between $S(E)$ datasets lead to significant changes in $\mathrm{SPR}_{w,\mathrm{Si}}(z)$. The percentage residuals presented in Fig.~\ref{fig5} were evaluated at each depth as the relative deviation between the converted PDD curve obtained using stopping-power dataset $k$ and the corresponding reference PDD curve:
\begin{equation}
\delta^{(k)}(z)
=
100\,
\frac{
\mathrm{PDD}^{(k)}(z)-\mathrm{PDD}_{\mathrm{ref}}(z)
}{
\mathrm{PDD}_{\mathrm{ref}}(z)
}.
\label{eq:pdd_residuals}
\end{equation}
In this expression, $\mathrm{PDD}^{(k)}(z)$ denotes the relative PDD calculated using dataset $k$, whereas $\mathrm{PDD}_{\mathrm{ref}}(z)$ represents the reference PDD curve reported by Faddegon \emph{et al}.~\cite{Faddegon:2015}.

Figure~\ref{fig5} also shows a systematic deviation outside the Bragg-peak region. From the first measurement point at a depth of $1.06$ mm, extending throughout most of the entrance and plateau region up to approximately $36$ mm, all converted relative PDD curves lie about $3$\% below the reference curve of Faddegon \emph{et al.}~\cite{Faddegon:2015}. In contrast to the dataset-dependent dispersion observed in the Bragg-peak region, the close agreement of the residuals in the plateau indicates the influence of the common high-energy treatment employed in the present stopping-power tables: for proton energies exceeding $10$ MeV, all datasets were extended using the SBETHE-corrected Bethe formalism. The systematic offset in the plateau therefore suggests that the high-energy representation of the water-to-silicon stopping-power ratio can introduce a consistent bias in the relative dose conversion, even in an energy domain where stopping-power calculations are typically considered well established.

Faddegon \emph{et al.}~\cite{Faddegon:2015} employed a mean excitation energy of $I = 78$ eV in their Monte Carlo simulations and observed that the calculated distal range underestimated the measured value. On the basis of this discrepancy, they inferred that the effective $I$ value of liquid water may be approximately $2-5$ eV higher than $78$ eV. The value $I = 79.4$ eV adopted in the present work, derived from an $f$-sum-rule-constrained experimental energy-loss function~\cite{MATIAS-PRL:2025}, is consistent with this indication. Nevertheless, variation of $I$ alone cannot be unambiguously identified as the sole cause of the approximately 3\% difference observed in the plateau region, because the present calculation also utilizes the SBETHE-corrected Bethe formalism, whereas the reference conversion relied on stopping-power ratios computed with GAMOS~\cite{GAMOS:2008}. A dedicated sensitivity study, in which the $I$ value and the high-energy stopping-power model are varied independently, would be necessary to quantify their respective contributions.

Although the depth-dose curves examined in this work represent relative PDD distributions rather than absolute absorbed-dose calibrations, a systematic difference of this magnitude in the plateau region remains metrologically significant. If an incorrect stopping-power ratio is used to convert the signal from a non-water-equivalent detector or calorimeter into absorbed dose to water, the resulting systematic bias could propagate into the beam calibration and the reported dose across the entrance and plateau regions, rather than being confined to the distal edge. Consequently, accurate high-energy stopping-power data are essential not only for range determination and Bragg-peak modeling but also for establishing traceable detector-to-water and calorimeter-to-water dose conversion coefficients~\cite{Palmans:2022,Lourenco:2022,Lourenco_2023,Palmans2013,Lourenco:2016}.

\subsection{Energy-dependent proton fluence spectra near the Bragg peak}

\begin{figure}[H]
\centering
\includegraphics[width=0.9\textwidth,clip]{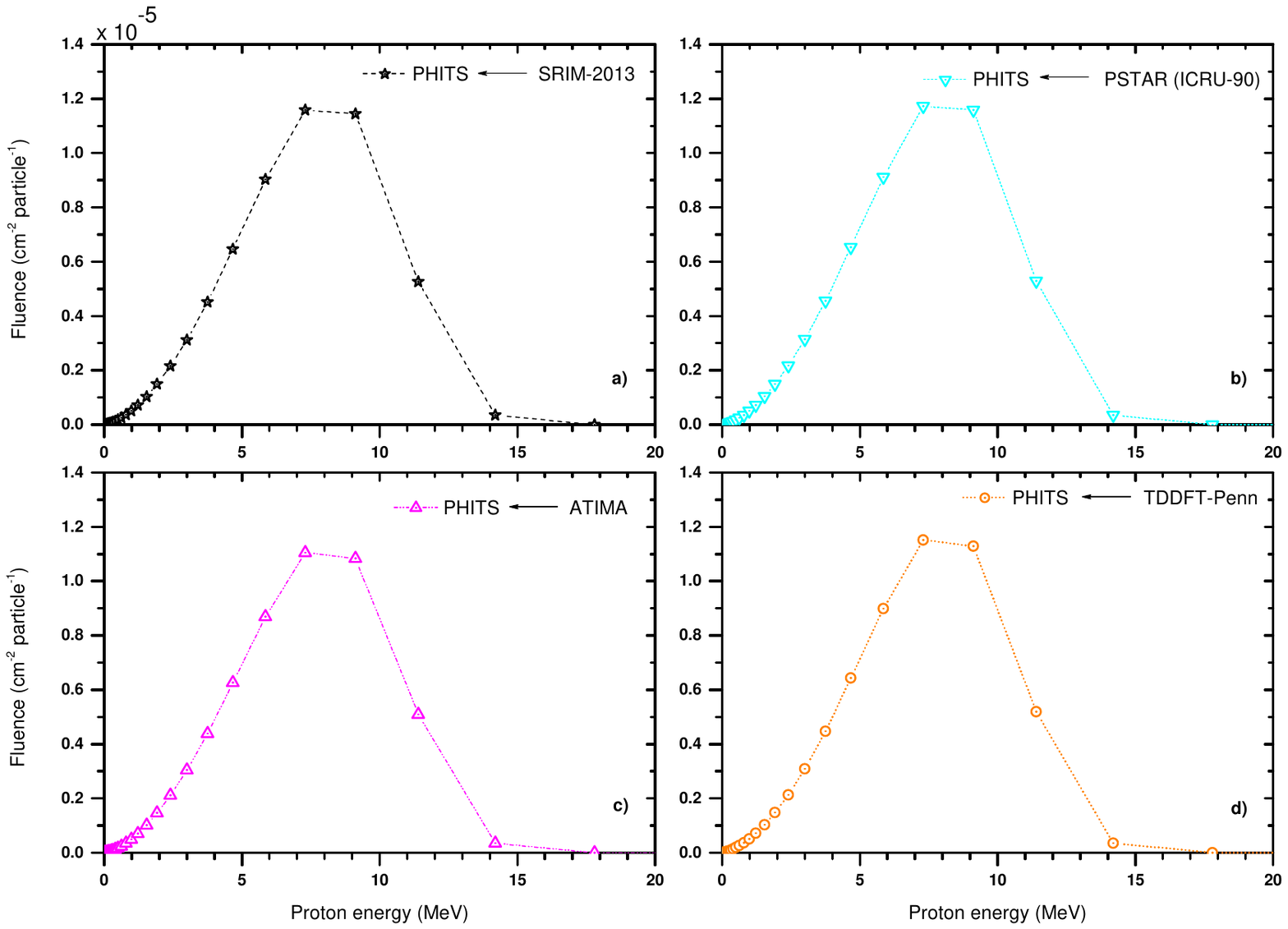}
\caption{Proton fluence spectra at $z=36.86$~mm, within the experimental Bragg-peak region, obtained in PHITS using different stopping-power datasets. Differences among spectra reflect dataset-dependent proton slowing upstream and in water, and help explain variations in the magnitude of the converted dose even when the Bragg-peak depth remains nearly unchanged.
}
\label{fig6}
\end{figure}

Figure~\ref{fig6} presents the energy-dependent proton fluence spectra evaluated within the scoring volume located at $z=36.86$ mm, corresponding to the maximum of the converted relative PDD curves. This depth is one of the two adjacent positions at which the original PTW PDI curve reaches $100$\% and lies $0.05$ mm distal to the experimentally defined midpoint of the Bragg-peak position of $36.81$ mm. Differences among the spectra reflect dataset-dependent proton energy loss in the upstream materials and in water. Because both the absorbed dose and the PDI-to-PDD conversion are spectrum-weighted quantities, changes in the low- to intermediate-energy component of the spectrum can modify the magnitude of the converted dose even when no dataset-dependent Bragg-peak displacement is observed.

\subsection{Bragg-peak depth from direct dose evaluation in water}

The depth-dose distributions computed directly in water were obtained using the PHITS code with each stopping-power dataset; the corresponding Bragg-peak depths and distal $80$\% ranges are summarized in Table~\ref{tab:range}. Across the PHITS configurations, the Bragg-peak depth ranged from $36.575$ to $36.675$ mm, corresponding to a spread of $0.1$ mm. This indicates that the choice of stopping-power dataset alone can lead to submillimeter shifts in the Bragg-peak position for this pristine beam. Within this interval, the PHITS $\leftarrow$ TDDFT-Penn configuration predicts the deepest Bragg-peak location, whereas the PHITS $\leftarrow$ ATIMA configuration yields the shallowest.


\begin{table}[H]
\caption{Bragg-peak depth and range extracted from depth-dose curves computed directly in water for each MC configuration. The experimental reference values correspond to the PTW silicon-diode dataset provided by Faddegon and used in the present analysis. The experimental PDI curve has equal maximum values at $36.76$ and $36.86$ mm; therefore, the representative Bragg-peak depth was defined as their midpoint, $(36.81\pm0.15)$ mm. The experimental methodology is described in Ref.~\cite{Faddegon:2015}.}
\centering
\begin{tabular}{l r r}
\hline
Method/datasets & Bragg-peak depth (mm) & Distal 80\% range, $R_{80}$ (mm)\\
\hline
PHITS $\leftarrow$ SRIM-2013   & 36.625 & 36.919\\
PHITS $\leftarrow$ PSTAR (ICRU-90)  & 36.625 & 36.922\\
PHITS $\leftarrow$ ATIMA  & 36.575 & 36.868\\
PHITS $\leftarrow$ TDDFT-Penn  & 36.675 & 36.940\\
PTW diode (experiment\cite{Faddegon:2015}) & 36.81 $\pm$ 0.15 & 37.35 $\pm$ 0.15\\
\hline
\end{tabular}
\label{tab:range}
\end{table}

Relative to the representative experimental Bragg-peak depth of $(36.81\pm0.15)$ mm, the PHITS $\leftarrow$ TDDFT-Penn configuration predicts a peak depth of $36.675$ mm, corresponding to an absolute deviation of $0.135$ mm. This result lies within the quoted experimental uncertainty interval. The PHITS $\leftarrow$ SRIM-2013 and PHITS $\leftarrow$ PSTAR configurations both predict $36.625$ mm, corresponding to deviations of $0.185$ mm, whereas PHITS $\leftarrow$ ATIMA predicts $36.575$ mm, corresponding to a deviation of $0.235$ mm. Thus, among the evaluated stopping-power options, only the TDDFT-Penn prediction lies within the experimental uncertainty of $\pm0.15$ mm. The SRIM-2013, PSTAR, and ATIMA results are slightly shallower than the lower bound of the experimental uncertainty, although all deviations remain submillimetric.

The distal $80$\% range, $R_{80}$, was also relatively stable but not strictly invariant across the stopping-power datasets. As reported in Table~\ref{tab:range}, the $R_{80}$ values obtained with the PHITS configurations ranged from $36.868$ to $36.940$ mm, corresponding to an inter-dataset spread of $0.072$ mm. However, all calculated values were $0.410-0.482$ mm shallower than the experimental reference value of $37.35$ mm. Consequently, the systematic offset relative to the measurement is substantially larger than the variation introduced by the different stopping-power datasets. This observation indicates that modeling components common to all PHITS configurations, such as the beam-line geometry, source characterization, upstream material budget, or the shared high-energy stopping-power formalism, are likely to dominate the absolute $R_{80}$ discrepancy. Additional targeted sensitivity analyses will be required to disentangle and quantify the individual contributions of these factors. Therefore, the range sensitivity remains submillimetric for this benchmark, but it should not be described as negligible or invariant.



\section{Discussion}

This study quantified how stopping-power dataset choices propagate into two quantities of practical relevance for proton dosimetry: (i) the conversion of normalized silicon-detector PDI data into relative dose-to-water distributions through water-to-silicon stopping-power ratios, and (ii) the Bragg-peak depth and distal range obtained from dose distributions calculated directly in water. For the pristine $67.5$ MeV benchmark beam and irradiation geometry considered here~\cite{Faddegon:2015}, the original PTW PDI curve contains two adjacent experimental maxima at $36.76$ and $36.86$ mm. Their midpoint, $(36.81\pm0.15)$ mm, was adopted as the representative experimental Bragg-peak depth. After applying different water-to-silicon stopping-power ratios, all converted relative PDD curves reached their maximum at $36.86$ mm. Therefore, the conversion did not introduce a dataset-dependent displacement of the Bragg peak; the $0.05$ mm difference relative to the midpoint-defined experimental reference is smaller than both the experimental sampling interval and the quoted depth uncertainty.

In contrast, the magnitude of the converted dose in the distal region showed a clear dependence on the selected stopping-power dataset. The depth-dependent water-to-silicon stopping-power ratio, $\mathrm{SPR}_{w,\mathrm{Si}}(z)$, varied substantially from entrance to distal depths, with relative changes ranging from $17.40$\% to $22.10$\% among the evaluated options. These differences led to dataset-dependent variations in the reconstructed dose near the Bragg peak, with deviations of up to $2.1$\%. The physical origin of this behavior is the increased sensitivity of the distal region to the low- to intermediate-energy part of the proton spectrum. In this energy interval, approximately $2-10$ MeV for the present benchmark, relatively small differences among stopping-power datasets can produce appreciable changes in $\mathrm{SPR}_{w,\mathrm{Si}}(z)$.

The energy-dependent proton fluence spectra calculated near the Bragg peak further support this interpretation. Although the macroscopic Bragg-peak position remains nearly unchanged after the silicon-to-water conversion, the dataset-specific description of proton slowing down in the upstream materials and in water modifies the local proton fluence distribution. Since both absorbed dose and stopping-power ratio corrections are spectrum-weighted quantities, changes in the spectral component sampled near the distal edge can alter the deposited dose per unit incident fluence and, consequently, the magnitude of the converted dose.

An additional observation is a systematic deviation spanning the entrance and plateau regions. From depths of $1.06$ mm to approximately $36$ mm, the relative PDD curves derived from the present $\mathrm{SPR}_{w,\mathrm{Si}}(z)$ calculations are systematically lower by about $3$\% relative to the reference curve, while the dataset-dependent discrepancy near the Bragg peak is approximately $2.1$\%. The near-identical plateau residuals obtained for the various datasets are consistent with their shared SBETHE-based high-energy continuation above $10$ MeV. This finding indicates that the accuracy of stopping-power data remains critical along the high-energy segment of the proton trajectory and is not confined to the distal-edge domain. Specifically, detector-to-water and calorimeter-to-water conversion procedures depend on stopping-power ratios and associated fluence and beam-quality correction factors; thus, a systematic bias in the entrance region can propagate into the reference-dose calibration and, consequently, into the normalization of the entire depth–dose distribution~\cite{MedinAndreo:1997,Palmans2013,Lourenco:2016,Palmans:2022,Lourenco:2022,Lourenco_2023}. However, because both the adopted mean excitation energy $I$ and the high-energy stopping-power formalism differ from those employed in the reference conversion, the present analysis does not allow for a clear separation of their respective contributions to the observed discrepancies.

For direct dose-to-water computations, the Bragg-peak depth exhibited a submillimetric yet measurable dependence on the chosen stopping-power dataset. Among the different PHITS setups, the Bragg-peak depth ranged from $36.575$ to $36.675$ mm, yielding a total spread of $0.1$ mm. Compared with the representative experimental depth of $36.81$ mm, the absolute deviations are $0.135$ mm for TDDFT-Penn, $0.185$ mm for both SRIM-2013 and PSTAR, and $0.235$ mm for ATIMA. Consequently, only the TDDFT-Penn result lies within the quoted experimental uncertainty of $\pm0.15$ mm. Thus, although the dataset dependence of the Bragg-peak depth is smaller than the corresponding effect on the converted dose magnitude, it is not negligible for high-precision benchmarking or for studies aimed at isolating range-uncertainty contributions. The distal $80$\% range, \(R_{80}\), likewise exhibited submillimetric variability across the Monte Carlo configurations, indicating a relatively small yet non-negligible dependence on the underlying dataset.

From a metrological perspective, these findings show that the stopping-power dataset used for measurement-to-dose conversion is a traceability-relevant source of systematic uncertainty. This is particularly important when silicon-detector measurements are used for Monte Carlo validation, inter-code comparisons, or the derivation of correction factors for non-water-equivalent detector materials. Whenever possible, the stopping-power data used in the detector-to-water conversion should be harmonized with those used in the reference Monte Carlo simulations. When such harmonization is not feasible, the dataset dependence should be explicitly included as a systematic component of the uncertainty budget with attention to the entrance, plateau, and distal regions of the depth-dose distribution.

The present study was intentionally restricted to a controlled benchmark based on a single pristine proton beam and a water-silicon conversion problem. Therefore, the conclusions should not be extrapolated directly to modulated clinical fields. In SOBP deliveries, multiple energy layers with different spectral hardness contribute to the composite dose distribution. Dataset-dependent differences in stopping power and in the transported proton fluence spectra may therefore accumulate differently from the pristine-beam case. Future work should extend the present analysis to SOBP fields, quantify the influence of upstream materials and transport settings on spectral differences, and assess whether restricted stopping powers and harmonized electromagnetic physics settings can reduce residual inter-code variability in distal dose magnitude.

\section{Conclusion}
In conclusion, the chosen stopping-power formalism systematically influences the silicon-to-water dose conversion along the entire proton trajectory. For the benchmark scenario investigated in this work, the converted relative PDD curves were approximately $3$\% lower than the reference curve in the entrance and plateau regions, while dataset-dependent deviations of up to $2.1$\% were observed near the Bragg peak. In direct dose-to-water calculations, the Bragg-peak position differed by $0.10$ mm and $R_{80}$ by $0.072$ mm among the various PHITS configurations. Relative to the PTW experimental Bragg-peak depth of $(36.81\pm0.15)$ mm, only the TDDFT-Penn prediction was within the quoted experimental uncertainty. Because the converted PDD curves represent relative dose distributions, these discrepancies should be interpreted as systematic biases in the detector-to-water conversion rather than as absolute-dose errors. These results underscore the importance of explicitly documenting and validating the stopping-power formulation, the selected mean excitation energies, and the normalization procedure employed in proton dosimetry. In situations where stopping-power implementations cannot be fully harmonized, their impact should be explicitly incorporated into the uncertainty budgets for proton-beam calibration, commissioning, quality assurance, and Monte Carlo-based verification.

\ack{We gratefully acknowledge Dr. Bruce A. Faddegon (University of California, San Francisco) for generously providing the original, unpublished, normalized PTW silicon-diode percentage-depth ionization (PDI) data that underlie the percentage-depth-dose curves reported in Ref.~\cite{Faddegon:2015}. Access to these data was indispensable for performing the silicon-to-water dose-conversion analysis presented in this work and, consequently, for the realization of the present study. This work has been done as a part of the Project INCT-Física Nuclear e Aplicações, Projeto No. 408419/2024-5; CNPq Projects Nos. 406982/2021-0 and 403722/2023-3; Spanish Ministerio de Ciencia e Innovación Project No. PID2021-122866NB-I00 financed by MCIN/AEI/10.13039/501100011033/ and ERDF A way of making Europe; Fundación Séneca-Agencia de Ciencia y Tecnología de la Región de Murcia Project No. 22081/PI/22; and FAPESP computer cluster process No. 2012/04583-8 and No. 2020/04867-2.}
\noindent


\roles{F. Matias: Conceptualization, Methodology, Software, Validation, Writing - Original Draft. J. M. B. Shorto: Funding acquisition, Data Curation, Writing - Review \& Editing. P. de Vera: Methodology, Validation, Writing - Review \& Editing. R. Garcia-Molina: Methodology, Validation, Writing - Review \& Editing. I. Abril: Methodology, Validation, Writing - Review \& Editing. T. F. Silva: Methodology, Validation, Writing - Review \& Editing. J. Pereira: Methodology, Validation, Writing - Review \& Editing. H. Yoriyaz: Methodology, Software, Validation, Writing - Original Draft. All authors reviewed the manuscript.}

\data{The PHITS input files and all derived datasets generated in this study are available from the corresponding author upon reasonable request. The original, unpublished PTW silicon-diode PDI dataset was provided by Dr. Bruce A. Faddegon and is not publicly accessible; access to these data is contingent upon obtaining permission from the data owner.}
\noindent



\bibliographystyle{iopart-num}   
\bibliography{references}

\end{document}